      [1999/12/01 v1.4c Il Nuovo Cimento]
\documentclass{cimento}
\title{Violation of the Feynman scaling law as a manifestation 
of non\-extensivity} 
\author{F.S.~Navarra\from{ins:SP}\thanks{e-mail: navarra@if.usp.br}\ETC,
O.V.~Utyuzh\from{ins:W}\thanks{e-mail: utyuzh@fuw.edu.pl},
G.~Wilk\from{ins:W}\thanks{e-mail: wilk@fuw.edu.pl}
        \atque
\\
Z.~W\l odarczyk\from{ins:K}\thanks{e-mail: wlod@pu.kielce.pl}}
\instlist{
  \inst{ins:SP} Instituto de F\'{\i}sica, Universidade de S\~{a}o Paulo,
                C.P. 66318,  05315-970 S\~{a}o Paulo, SP, Brazil
  \inst{ins:W} The Andrzej Soltan Institute for Nuclear Studies,
               ul. Ho\.za 69, 00-681 Warsaw, Poland  
  \inst{ins:K} Institute of Physics, Pedagogical University,
               ul. Konopnickiej 15, 25-405 Kielce, Poland 
               }
\PACSes{\PACSit{96.40}{Cosmic rays}
\PACSit{01.30.Cc}{Conference proceedings}}
\begin{document}

\maketitle

\begin{abstract}
We demonstrate that the apparently {\it ad hoc} parametrization
of the particle production spectra discussed in the literature
and used in description of cosmic ray data can be {\it derived}
from the information theory approach to multiparticle production
processes. In particular, the violation of the Feynman scaling law
can be interpreted as a manifestation of nonextensivity of the
production processes.
\end{abstract}

\section{Introduction}

The shape of the $x=E/E_0$ spectra of secondaries is of great
importance in all investigations concerning developments of cosmic ray
cascades \cite{ref:AO} (cf. also \cite{ref:WdWo}). The crucial problem of
practical importance is the {\it existence} or {\it nonexistence} of
the Feynman scalling, which says that $x$-spectra of secondaries are
energy independent. Because in cosmic ray applications one is
sensitive essentially to large $x$ region (of, say, $x > 0.1$), the
commonly used formula \cite{ref:AO} 
\begin{equation}
\frac{dN}{dx}\, =\, Da\frac{\left(1\, - a' x\right)^4}{x} \label{eq:AO}
\end{equation}
stresses this fact by power-like form with exponent obtained from 
the $x \rightarrow 1$ limit of fragmentation data \cite{ref:AO}. 
Here $a'$ is parameter responsible for the Feynman
scaling violation whereas $D$ and $a$ are additional parameters obtained from
the fit to data - all are energy dependent (cf. \cite{ref:AO} for
details). 

We shall demonstrate that (\ref{eq:AO}) emerges in 
natural way from the information theory approach to hadronization
processes presented in \cite{ref:MAXENT} and extended to allow for the
possible nonextensivity of the hadronization process by applying
Tsallis entropy \cite{ref:T}. (See  \cite{ref:T-rev} for details
of notion of nonextensivity and \cite{ref:DENTON} for its application
to high energy reactions.).

\section{Particle production spectra from information theory}

In order to describe the hadronization process in information 
theory one uses  \cite{ref:MAXENT} the  
{\it least biased and most plausible} single
particle distribution $f(y) = \frac{1}{N} \frac{dN}{dy}$ (where $y$
is rapidity) resulting from hadronization process in which a mass
$M$ hadronizes into $N$ secondaries of mean transverse mass $\mu_T =
\sqrt{\mu^2 + \langle p_T \rangle^2}$ each (for simplicity
we consider only one-dimensional hadronization with limited
transverse momenta which can, however, depend on $M$ and $N$).
This is done maximalizing the Shannon (or Boltzmann-Gibbs) information
entropy, $S\, =\, -\, \int dy\, f(y)\, \ln f(y)$, under constraints of
normalization ( $\int dy f(y) = 1$ ) and energy conservation ($\int
dy\mu_T \cosh y f(y) = M/N$ ), which leads to the following (extensive)
distribution function cf. \cite{ref:MAXENT} for details):
\begin{equation}
f(y)\, =\, \frac{1}{Z(M,N)} \exp\left[ - \beta(M,N)\cdot \mu_T \cosh y
\right] . \label{eq:MAXENT}
\end{equation}
Here $Z(M,N)$ comes from the normalization of $f(y)$ whereas the
Lagrange multiplier $\beta(M,N)$ is to be calculated from the energy 
conservation constraint. Notice that {\it there is no free parameter
here}. This should be contrasted with the popular use of eq.
(\ref{eq:MAXENT}) as a "thermodynamical parametrizations" with
inverse "temperature" $1/\beta = T$ beeing a free, positively defined
parameter. In \cite{ref:MAXENT} it was shown that, in a wide range of
energies and multiplicities, $\beta (M,N) \sim \frac{N}{M} =
\frac{1}{\sqrt{s}}\frac{N}{K}$. 

However, the comparison with data cannot be done in such model 
independent way because of the fluctuations of inelasticity $K$
(given by inelasticity distribution $\chi(K)$ \cite{ref:INEL})
causing fluctuations of $M$ and because of fluctuations of number of
particles $N=N(M)$ produced from a mass $M$ given by multiparticle 
distribution $P(N;M)$ \cite{ref:MULTI} (its actual shape accounts here
for the possible multisource structure of the production process).
One has therefore 
\begin{equation}
\frac{dN}{dy}\, =\, \int^1_0 dK \chi(K)\, \sum_N P(N;K\sqrt{s}) 
                N\frac{1}{Z(K\sqrt{s},N)}\exp\left[ -
                \beta(K\sqrt{s};N) \mu_T \cosh y \right], \label{eq:FULL}
\end{equation}
where $\beta(K\sqrt{s},N)$ is still calculated from the energy
conservation constraint but now for a given event, i.e., it is
a fluctuating quantity with fluctuations resembling gamma-like
distribution (and characterised by mean value $\bar{\beta}$ and 
normalised variance $\omega_{\beta}$).

It has been shown in \cite{ref:PRL,ref:DENTON} that such fluctuations 
of the parameter of the exponential distribution convert it to a
power-like distribution:
$\exp( - \hat{x}/x_0) \rightarrow
     \exp_q( - \hat{x}/x_0) =
     ( 1 - \hat{x}/x_0/\alpha)^{\alpha}$ with $\alpha =\pm
1/\omega_{\beta}$ in our case where
$\hat{x}=\frac{2\mu_T}{\sqrt{s}}\cosh y$ and $x_0$ is a parameter
connected with the mean value of $\beta \sim \frac{N}{K}$. 

This is precisely the form corresponding to distribution obtained
using non-extensive Tsallis $q$-entropy \cite{ref:T,ref:T-rev},
instead of Shannon one (and reproducing it in the limit 
$q\rightarrow 1$),
$S_q = - (1 - \int dy\, [f(y)]^q)/(1 - q)
\rightarrow 
S_{q=1} = - \int dy f(y) \ln f(y)$,
together with the modified constraint equation $\int dy \mu_T \cosh y
[f(y)]^q = M/N$. Using it one gets instead of (\ref{eq:MAXENT}) its
nonextensive version: 
\begin{equation}
f_q(y)\, =\, \frac{1}{Z_q(M,N)} \left[ 1\, -\, (1 - q)
                                   \beta_q(M,N)\cdot \mu_T  
                \cosh y \right]^{\frac{1}{1-q}} . \label{eq:MAXENTq}
\end{equation}
Nonextensivity means that the  entropy of the composition
$(A+B)$ of two independent systems $A$ and $B$ is equal to
\cite{ref:T-rev} $S^{(A+B)}_q\, =\, S^{(A)}_q\, +\, S^{(B)}_q\, +
               \, (1 - q)\, S^{(A)}_q\cdot S^{(B)}_q$
proceeding to its usual additive form only in the $q=1$ limit. It
arises whenever in the system one enconters long-range correlations,
memory effects or fractal structure of the corresponding space-time
or phase space. Such situation is expected to occur also in the
hadronization processes \cite{ref:DENTON}.

\section{Comparison with experimental data}

\begin{figure}[h]
\setlength{\unitlength}{1cm}
\begin{picture}(10.,11.)
\includegraphics{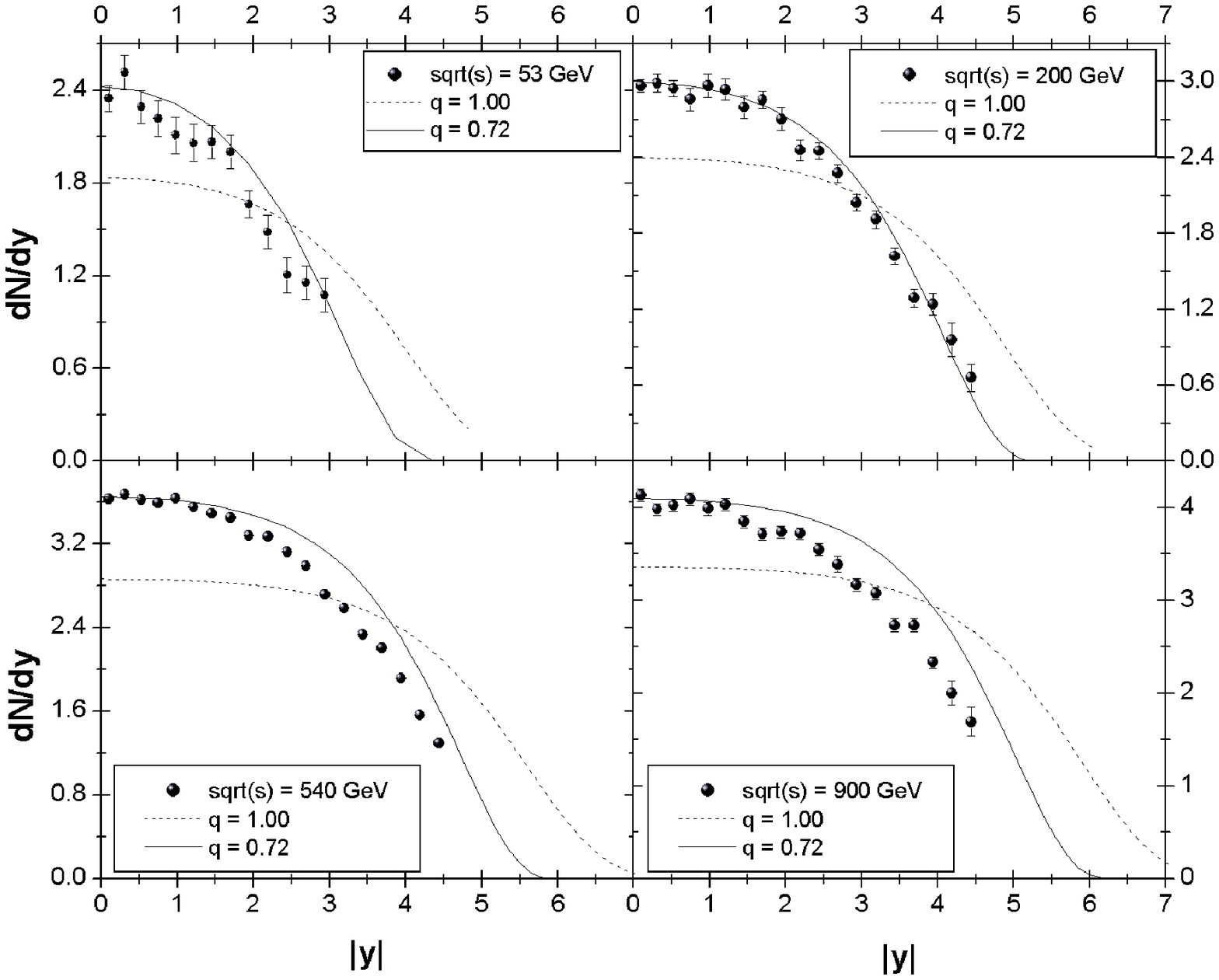}
\end{picture}
\vspace{-0.5cm}
\end{figure}
\vspace{-0.5cm}
\begin{minipage}[h]{14cm}
\noindent
Fig. 1. Comparison of eq. (\ref{eq:FORM}) with UA5 \cite{ref:UA5} data for
$q=1$ and $q=072$. 
\end{minipage}
\vspace{5mm}

We make therefore the following conjecture: the convolution
(\ref{eq:FULL}) can, in practice, be replaced by the following simple
one parameter formula of the type of eq. (\ref{eq:AO}) which should
be used to describe the existing data of \cite{ref:UA5,ref:P238,ref:UA7}: 
\begin{equation}
\frac{dN}{dy}\, =\, \langle N(s)\rangle \frac{1}{Z_q}\, \left[1\, -\, (1-q)
\beta_q(\sqrt{s},\frac{3}{2}\langle N(s)\rangle) \mu_T \cosh y \right]^{\frac{1}{1-q}}.
\label{eq:FORM}
\end{equation}
The only free parameter (characterising the strenght of fluctuations
in the system according to the previous discussion) is the
nonextensivity parameter $q$. Here $Z_q = \int dy \exp_q (- \beta_q
\mu_T \cosh y)$ and $\langle N(s)\rangle$ is the mean (single
non-diffractive) charged multiplicity at given energy $\sqrt{s}$.
Notice that because $\beta_q$ is calculated from the energy
conservation constraint which involves all produced particles (i.e.,
total multiplicity) it is calculated here for $\frac{3}{2}\langle
N(s)\rangle = \langle N_{total}(s)\rangle$ particles. For the same
reason care must be taken when one addresses data at $\sqrt{s} = 630$
GeV as part of them is for charged and part for neutral particles
only. In both cases the $\beta_q$ must be the same (calculated for
total multiplicity at given energy) whereas multiplicity in front of
the formula has to be chosen accordingly to 
the actual situation. 

In Fig. 1 we show our results both for $q=1$ and our best fit with
$q=0.72$. In all calculations the experimentally observed
variation of $\mu_T$ with energy has also been accounted for by using
the following simple interpolation formula: $\mu_T = 0.3 +
0.044\ln(\sqrt{s}/20.)$ GeV. The results are reasonable, especially
for $53$ and $200$ GeV. For higher energies our distributions start
to be broader than data and this cannot be improved by changing $q$
as diminishing its value in order to make distributions narrower will
spoil the agreement with data for small rapidities. It turns out that UA7
and P238 data cannot be fitted together with UA5 data, as they demand
a slightly bigger value of $q=0.85$, cf. Fig. 2. Notice that P238
data are for charged and UA7 data for neutral particles therefore
they must be described with, respectively, $\frac{2}{3}\langle
N_{total}\rangle$ and $\frac{1}{3}\langle N_{total}\rangle$ in eq. (\ref{eq:FORM}),
this leads to differences clearly seen in Fig. 2.

\begin{figure}[h]
\unitlength1cm
\begin{minipage}[t]{6.5cm}
\begin{picture}(6.5,5.5)
\includegraphics{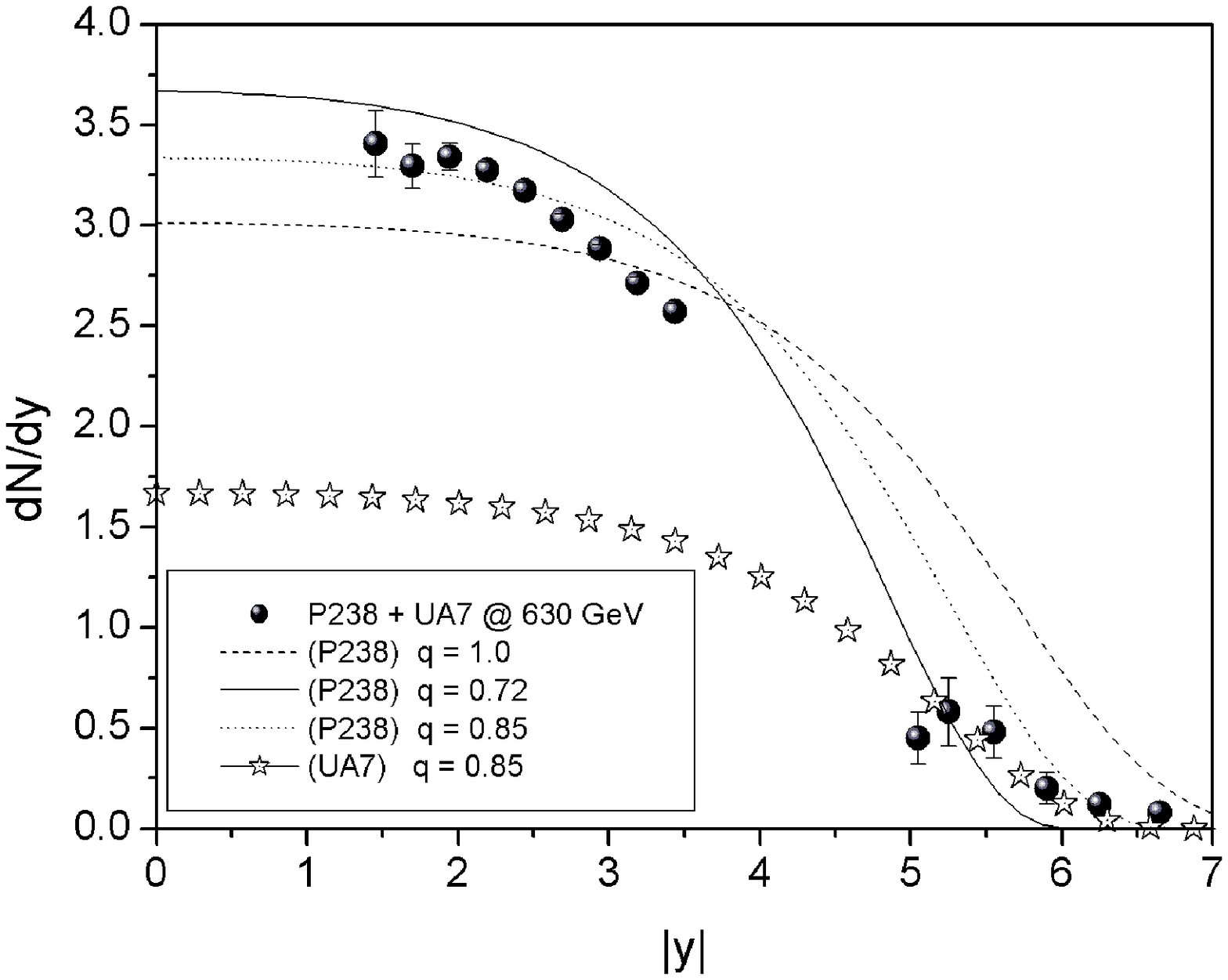}
\end{picture}
\end{minipage}
\hfill
\begin{minipage}[t]{6.5cm}
\begin{picture}(6.5,5.5)
\includegraphics{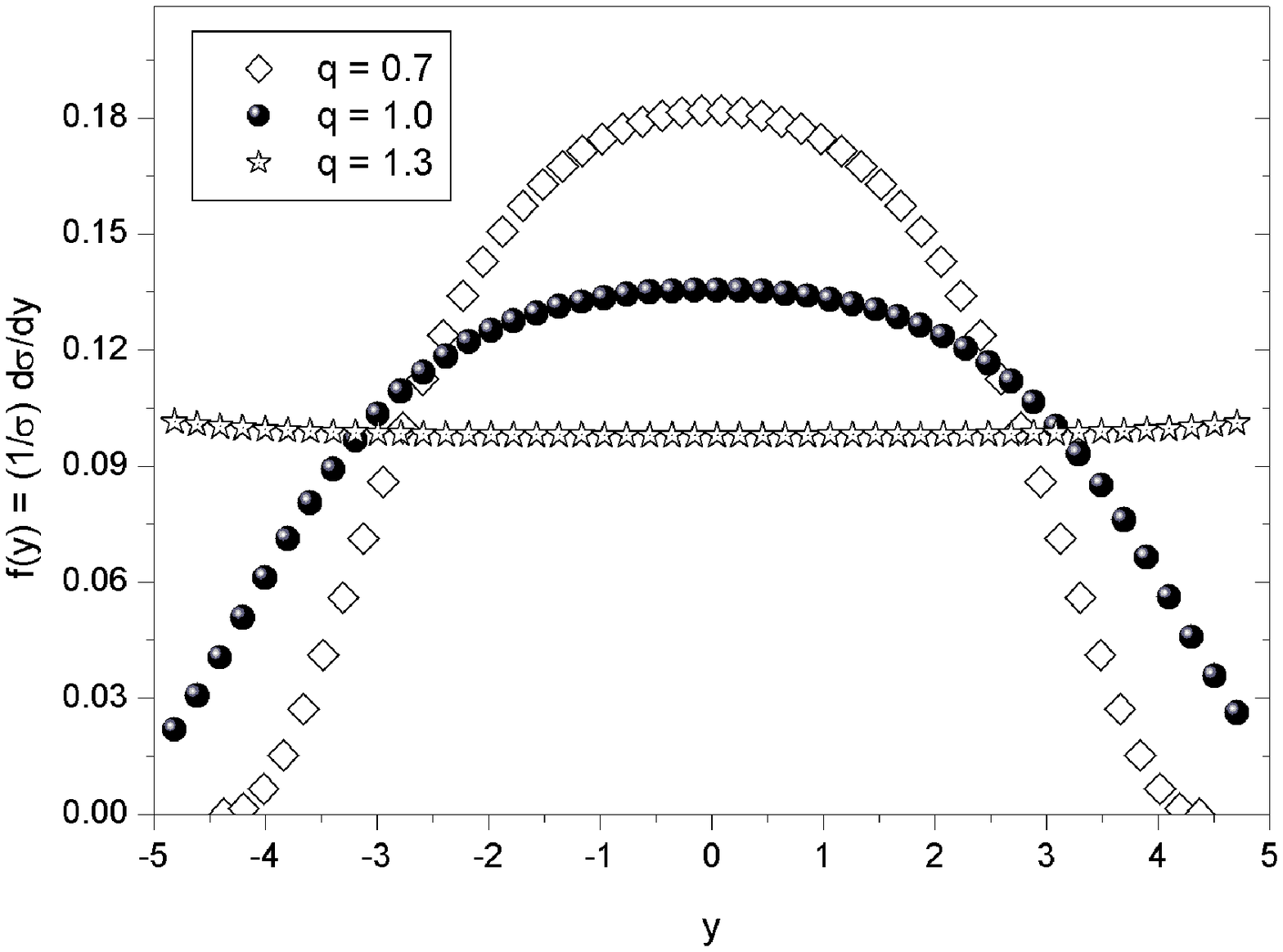}
\end{picture}
\end{minipage}
\end{figure}
\begin{figure}[h]
\unitlength1cm
\begin{minipage}[t]{6.5cm}
\begin{picture}(6.5,3)
\includegraphics{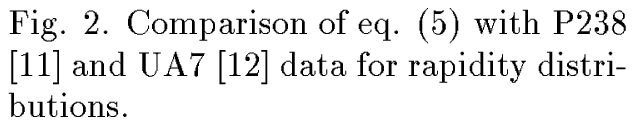}
\end{picture}
\end{minipage}
\hfill
\begin{minipage}[t]{6.5cm}
\begin{picture}(6.5,3)
\includegraphics{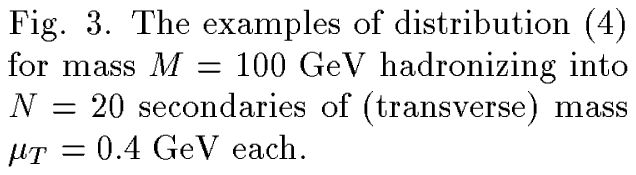}
\end{picture}
\end{minipage}
\vspace{-2cm}
\end{figure}

\section{Summary and conclusions}

Our approach consists in noticing the striking similarity of eq.
(\ref{eq:MAXENTq}) to the eq. (\ref{eq:AO}) of \cite{ref:AO} ($dy =
dx/x$). Fig. 3 shows the characteristic features of $f_q(y)$ for the
case of hadronization of fireball of mass $M=100$ GeV into $N=20$
particles of transverse mass $\mu_T=0.4$ GeV. Notice that for $q<1$
one obtains an  increase of particle densities in the central region
connected with its decrease at the edges of rapidity range -
clearly  resembling the characteristic pattern of Feynman scaling
violation.  We have replaced therefore the "exact"
formula, (\ref{eq:FULL}) by a one parameter fit represented by
(\ref{eq:FORM}) where $q$ summarizes the action of the averaging over
the fluctuations caused by the initial conditions of reaction
represented by the inelasticity and multiplicity distributions. 
We find it very amazing that such simple approach coincides
practically with empirical formula (\ref{eq:AO}) and with only one
parameter $q$ describes fairly well all data. Notice that $q=0.72$ is
not very far from $q=0.75$, which gives power $1/(1-q)=4$ in (\ref{eq:AO}).
Notice also that our general formula allows for description of
small $x$ region as well. It would certainly be interesting to connect
$q$ directly to the parameters describing inelasticity and
multiplicity distributions or to descriptions of leading particles,
for example of the type of that presented in \cite{ref:FN}. Our results
seem to indicate that most probably parameter $q$ should be
$x$-dependent reflecting different character of fluctuations 
of the quantity $N/M$ \cite{ref:PRL} in the 
central (mostly pionization) and fragmentation regions. This problem
will be addressed elsewhere.

\acknowledgments
Authors ZW and GW are gratefull to the organizers of the {\it
Chacaltaya Meeting on Cosmic Ray Physics} for their support and
hospitality. The partial support of Polish Committee for 
Scientific Research (grants 2P03B 011 18 and 
621/E-78/SPUB/CERN/P-03/DZ4/99) is acknowledged.

\end{document}